# LLRF CONTROLS IN SUPERKEKB PHASE-1 COMMISSIONING

T. Kobayashi[†], K. Akai, K. Ebihara, A. Kabe, K. Nakanishi, M. Nishiwaki, J. Odagiri, S. Yoshimoto, KEK, Tsukuba, Ibaraki, Japan
K. Hirosawa, SOKENDAI, Tsukuba, Ibaraki, Japan

*Abstract*

First beam commissioning of SuperKEKB (Phase-1), which is an asymmetry double ring collider of 7-GeV electron and 4-GeV positron beams, which had started from February, has been successfully accomplished at the end of June 2016, and the desired beam current for Phase-1 was achieved in both rings. This paper summarize the operation results related to low level RF (LLRF) control issues during the Phase-1 commissioning, including the system tuning, the coupled bunch instability and the bunch gap transient effect. RF system of SuperKEKB consists of about thirty klystron stations in both rings. Newly developed LLRF control systems were applied to the nine stations among the thirty for Phase-1. The RF reference signal distribution system has been also upgraded for SuperKEKB. These new systems worked well without serious problem and they contributed to smooth progress of the commissioning. The old existing systems, which had been used in the KEKB operation, were still reused for the most stations, and they also worked as soundly as performed in the KEKB operation.

## INTRODUCTION

KEKB is an asymmetric energy collider consisting of an 8 GeV electron ring (high-energy ring, HER) and a 3.5 GeV positron ring (low-energy ring, LER), which was operated from 1998 to 2010 [1]. It obtained the world record in luminosity of $2.11 \times 10^{34}$ cm$^{-2}$s$^{-1}$. To increase the luminosity, a high-current beam is needed in both rings. One serious concern for high-current storage rings is the coupled-bunch instability caused by the accelerating mode of the cavities. This issue arises from the large detuning of the resonant frequency of the cavities that is needed to compensate for the reactive component of the beam loading [2]. Two types of cavities that mitigate this problem are used in KEKB [3, 4]: one is the ARES normal conducting three-cavity system [5, 6] and the other is the superconducting cavity (SCC) [7, 8]. The detuning frequency of these cavities is reduced owing to the high stored energy in these cavities.

The ARES is a unique cavity, which is specialized for KEKB. It consists of a three-cavity system operated in the π/2 mode: the accelerating (A-) cavity is coupled to a storage (S-) cavity via a coupling (C-) cavity as shown in Fig. 1 [9]. The A-cavity is structured to damp higher-order modes (HOM). The C-cavity is equipped with a to damp parasitic 0 and π-modes. The π/2 mode has a high

___
\* The authors of this work grant the arXiv.org and LLRF Workshop's International Organizing Committee a non-exclusive and irrevocable license to distribute the article, and certify that they have the right to grant this license.
† tetsuya.kobayashi@kek.jp

Q-value even with a C-cavity with a very low Q-value of about 100. In LER, where a higher beam current is stored than in HER, only the ARES cavities were used. For details regarding the RF systems of KEKB, see Refs. [3, 4]. The RF issues to be considered for the heavy-beam current storage are summarized in Ref. [10].

KEKB is being upgraded to SuperKEKB, which is aiming at a 40 times higher luminosity than KEKB [11, 12]. The RF related machine parameters are shown in Table 1. The RF systems are being reinforced to handle twice as large stored beam currents in both rings and much higher beam power (compared to KEKB) [13]. ARES and SCC will be reused with the reinforcements. The RF power source systems, including klystrons, waveguides, and cooling systems, also need to be reinforced to increase the driving RF power to provide larger beam power. Furthermore, a new low-level RF (LLRF) control system, which is based on a recent digital control technique using field-programmable gate arrays (FPGAs), has been developed to realize higher accuracy and greater flexibility [14]. For nine RF stations, among a total of thirty, the LLRF control system used in KEKB has been replaced with new ones.

The first beam commissioning of SuperKEKB (Phase-1) was accomplished in 2016. The RF systems and the new LLRF control systems were soundly worked. The desired beam current of 1A for Phase-1 was successfully achieved and the vacuum scrubbing was sufficiently progressed.

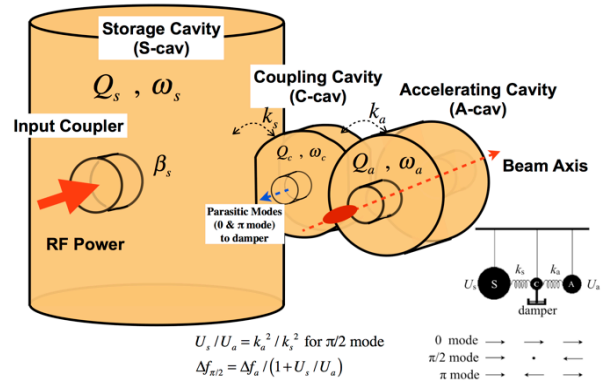

Figure 1: Illustration of the ARES cavity structure.

## RF SYSTEM ARRANGEMENT

RF related parameters of SuperKEKB are shown in Table 1 in contradistinction with those of KEKB. The RF system layout of KEKB is shown in Fig. 2. The planed arrangement for SuperKEKB at the design beam current is also shown in the figure (lower side). We have about thirty stations of the RF power source (klystron and the

LLRF control). SCC's are used in HER at 8 stations in Nikko section. For the other section, ARES cavity units are used. In LER, all cavities are the ARES type. In the KEKB operation, one klystron has driven two ARES units. On the other, in SuperKEKB, one klystron should drive only one ARES unit (so called "one-to-one configuration" is necessary), because the beam power per ARES will be three times higher than that of KEKB (see Table 1). The cavity input power will be about 750 kW (cavity wall loss + beam power) in SuperKEKB, while the maximum klystron output power is about 1MW. Accordingly additional klystrons are needed for the upgrade. The input coupler of ARES has been already reinforced for the increased input power.

For Phase 1 as a first step, six ARES units of OHO-D5 section were relocated from HER to LER, and the configuration of them was changed to the one-to-one configuration. Additionally the configuration of D7-C, D7-D, D8-C and D8-D stations was also changed to the one-to-one configuration for Phase-1.

Table 1: RF related parameters of KEK and SuperKEKB

| Parameter | unit | KEKB (achieved) | | SuperKEKB (design) | |
| --- | --- | --- | --- | --- | --- |
| Ring | | HER | LER | HER | LER |
| Energy | GeV | 8.0 | 3.5 | 7.0 | 4.0 |
| Beam Current | A | 1.4 | 2 | 2.6 | 3.6 |
| Number of Bunches | | 1585 | 1585 | 2500 | 2500 |
| Bunch Length | mm | 6-7 | 6-7 | 5 | 6 |
| Total Beam Power | MW | ~5.0 | ~3.5 | 8.0 | 8.3 |
| Total RF Voltage | MV | 15.0 | 8.0 | 15.8 | 9.4 |
| | | ARES SCC | ARES | ARES SCC | ARES |
| Number of Cavities | | 10  2 | 8   20 | 10   8 | 8   14 |
| Klystron : Cavity | | 1:2  1:1 | 1:1  1:2 | 1:1  1:1 | 1:2  1:1 |
| RF Voltage (Max.) | MV/cav. | 0.5  1.5 | 0.5 | 0.5  1.5 | 0.5 |
| Beam Power (Max.) | kW/cav. | 200  550 | 400  200 | 600  400 | 200  600 |

## LLRF CONTROL SYSTEM

Accuracy and flexibility in accelerating field control are very essential for storage of high-current and high-quality beam without instability. Therefore, new low level RF (LLRF) control system, which is based on recent digital architecture, was developed for the SuperKEKB. Figure 3 shows a picture of a mass-production model of the new LLRF system for the SuperKEKB. A block diagram of an ARES cavity driving system is shown Fig. 4. The principal functions of this system are performed by five FPGA boards which work on MicroTCA platform as advanced mezzanine cards [15]: Vc-feedback controller (FBCNT), cavity-tuner controller (TNRCNT), inter-lock handler (INTLCNT), RF-level detector for the interlock and arc-discharge photo-signal detector. As shown in Fig. 4, the new LLRF control system handles I/Q components of controlling signals in the FPGAs. For slow interlocks (e.g. vacuum, cooling water) and sequence control, a PLC is utilized. EPICS-IOC on Linux -OS is embedded in each of the FPGA boards and the PLC [16].

At 9 stations of Oho D4&D5 (6@D5 + 3@D4), the LLRF control systems were replaced with new digital control systems for Phase-1 as shown Fig. 5 and Fig. 6. All of new systems successfully worked well without problem. Some software bugs found during the operation were fixed. The DR-LLRF control system has already installed in DR control room. It is almost the same as MR one, except 3-cavity vector-sum control is needed. In the present stage, the number of cavities is two.

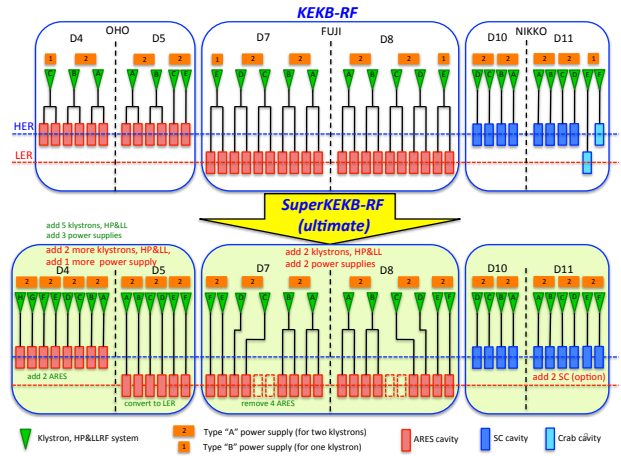

Figure 2: RF system arrangement of KEKB and plan for SuperKEKB ultimate stage.

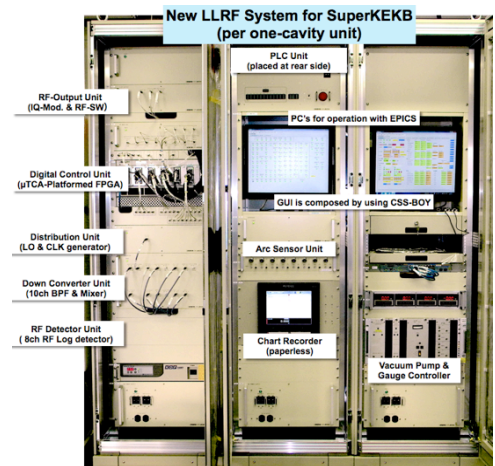

Figure 3: LLRF control system for SuperKEKB.

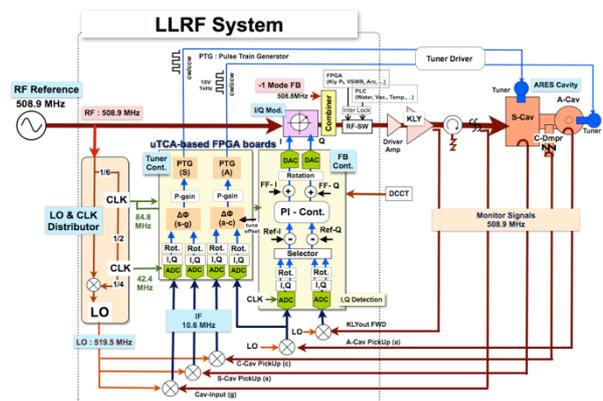

Figure 4: Block diagram for ARES cavity control.

On the other hands, the other stations were still operated with existing (old analogue) LLRF control systems, which had been used in the KEKB operation. These systems are composed of combination of NIM standard analogue modules as shown in Fig. 7. They are controlled remotely via CAMAC system. All systems also soundly worked as well as operated in the KEKB operation, although many old defective modules were replaced with spares in the maintenance works.

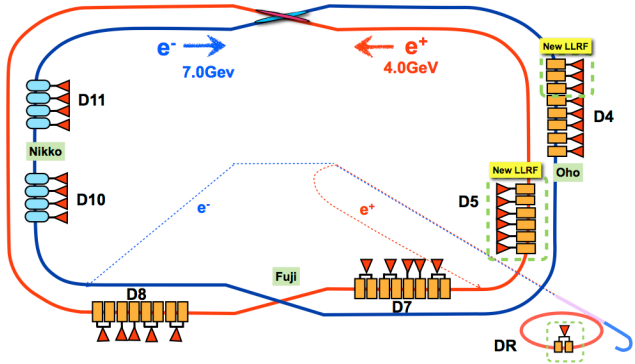

Figure 5: RF system layout for the Phase-1. Nine LLRF stations were replaced with the new ones. DR-LLRF control system was also newly installed.

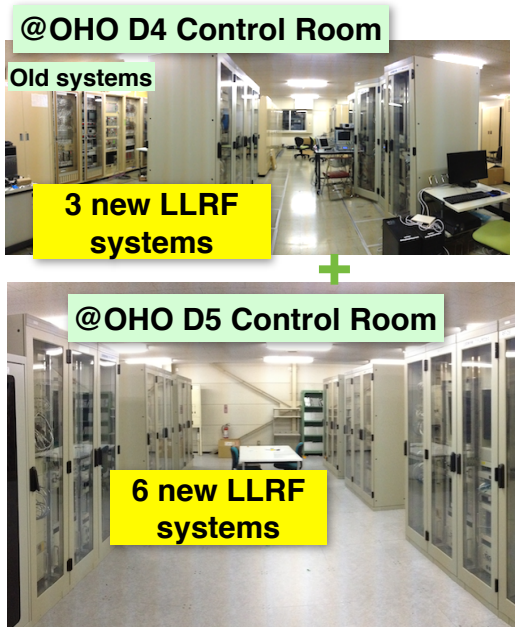

Figure 6: Installation appearance of the new LLRF control systems in the RF control rooms at D4 (upper) and D5 (lower) section.

RF reference distribution system was also upgrade for SuperKEKB [17]. RF reference signal is optically distributed into 8 sections by means of "Star" topology configuration from the central control room (CCR). "Phase Stabilized Optical Fiber", which has quite small thermal coefficient (< 1ppm/°C), is adapted. For the thermal phase drift compensation, optical delay line is controlled digitally at CCR for all transfer lines as shown in Fig 8. The short term stability (time jitter) is about 0.1 ps (rms), and the long term stability (pk-pk) is ±0.1° = ±0.55 ps at 508.9MHz (expected by the optical delay control).

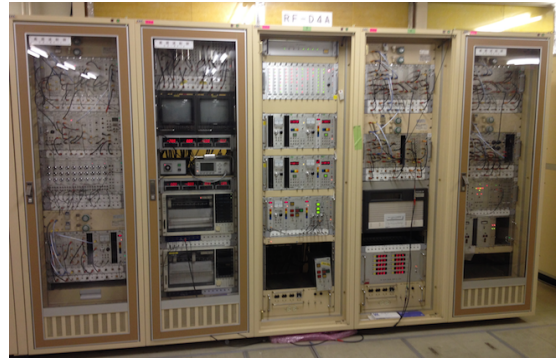

Figure 7: Old LLRF control system, which was used in KEKB operation, continues in use for SuperKEKB.

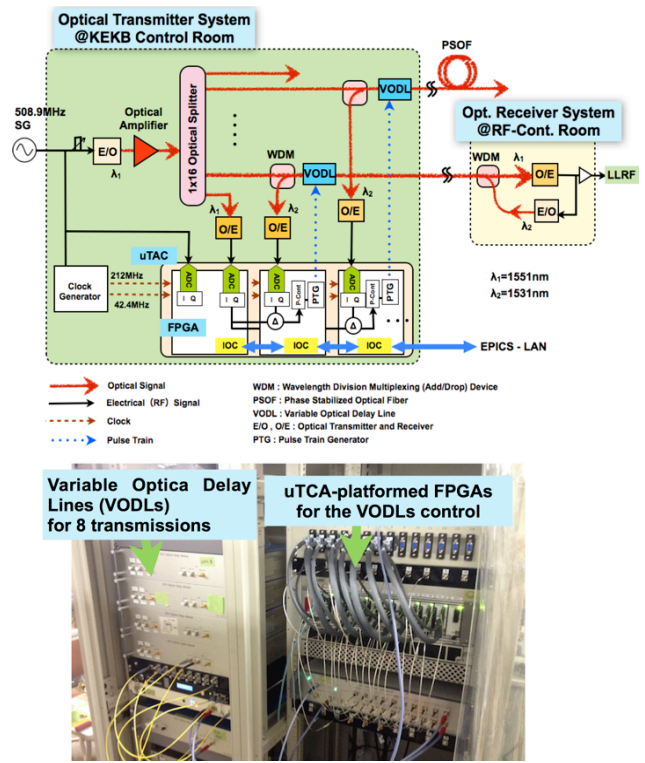

Figure 8: New RF reference signal distribution system for SuperKEKB. Block diagram of the reference distribution (upper side) and the photo of VODL control system (lower side) are shown.

The Phase-1 commissioning result is shown in Fig. 9. History of the stored beam current with beam dose (upper side) and the total acceleration voltage called total-Vc (lower side) of the both ring during Phase-1 is plotted in the figure. RF systems worked well without serious trouble. Target beam current of ~1A for Phase-1 was successfully achieved in both ring and vacuum scrubbing has been sufficiently progressed.

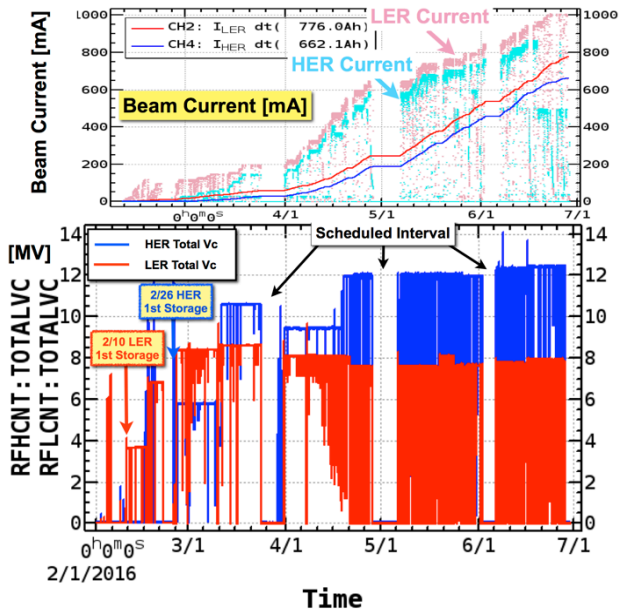

Figure 9: History of the stored beam current, the beam dose (upper side), and the total acceleration voltage called Total-Vc (lower side) for the both ring.

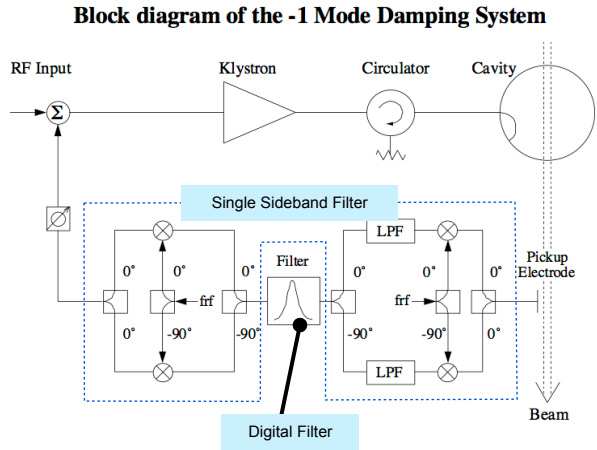

Figure 10: Block diagram of the $\mu$=-1 mode damping system, which had been used in KEKB operation. The $\mu$=-1 mode digital feedback selectively reduces impedance at the driving frequency.

## COUPLER BUNCH INSTABILITY DUE TO ACCELATING MODE

In HER, over the 400-mA beam current, the $\mu$=-1 mode instability due to the detuned cavities (parked with some reasons) was excited. It could not be suppressed by the tuner adjustment. Consequently, the $\mu$=-1 mode damper system, which had been used in KEKB operation as shown in Fig. 10, was applied to the D4 station. It worked well to suppress the $\mu$=-1 mode successfully as shown Fig. 11 and the beam current could be increased.

At the design beam current of SuperKEKB, the growth rate of the $\mu$=-2 mode instability will be close to the radiation damping rate. Therefore, the $\mu$=-2 mode damper system is additionally necessary for Phase-2. New damper system with new digital filters is now under development for Phase-2 [18]. It will be available for $\mu$=-2-1, -2 and -3 modes in parallel. Respective feedback phase for each mode can be adjusted independently in the digital filter.

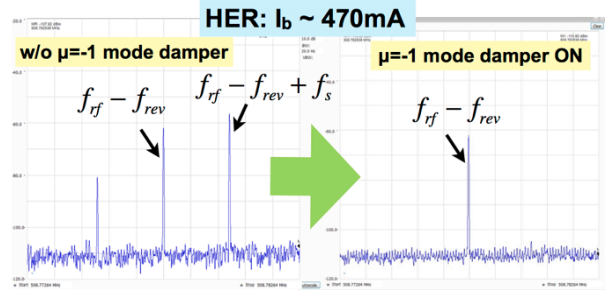

Figure 11: HER beam spectrum at the $\mu$=-1 mode. By applying the $\mu$=-1 mode damper, the couple bunch instability was successfully suppressed.

## BUNCH GAP TRANSIENT EFFECT ON BEAM PHASE

In generally, for a high-current multi-bunch storage ring, a bunch train has a gap of empty buckets in order to allow for the rise time of a beam abort kicker. The empty gap is also effective in clearing ions in electron storage rings. However, the gap modulates the amplitude and phase of the accelerating cavity field. Consequently, the longitudinal synchronous position is shifted bunch-by-bunch along the train, which shifts the collision point of each bunch.

The observed phase-shift due to the bunch gap effect, which was measured along the train in KEKB operation, agreed well with a simulation and a simple analytical formula in most part of the train. However, a rapid phase change was also observed at the leading part of the train, which was not predicted by the simulation or by the analytical form [3]. In order to understand the cause of this observation, we have developed an advanced simulation, which treats the transient loading in each of the three cavities of ARES [19]. The new simulation can reproduce well the observed rapid phase change. Accordingly, it was clarified that the rapid phase change at the leading part of the train is caused by a transient loading in the three-cavity system of ARES: the rapid phase change is attributed to the parasitic (0 & $\pi$) modes of ARES.

Figure 12 shows RF-phase transient of the accelerating cavity of ARES. It was measured by the new digital LLRF control system in the Phase-1 commissioning. The time interval of 10 microseconds is the revolution period. As shown in the figure, the phase modulation due to the bunch gap transient effect was clearly observed by the new LLRF system, and the rapid phase change caused by the parasitic mode of ARES can be also found at the leading part of the train. The simulation result (dashed red line) agrees well with the measurement. In the simulation, function of feedback control for cavity voltage regulation by LLRF control system is also included. From these results, validity of the new simulation code can evidently confirmed.

From the new simulation study, it is predicted that the rapid phase change caused by bunch gap will be about 6 degrees in cavity field at the design beam current of SuperKEKB. The collision point shift (relative phase difference between the two rings) due to such large phase change could make significant reduction of the luminosity in future collision experiment. The measures to avoid the luminosity reduction have been also proposed by the simulation study [19].

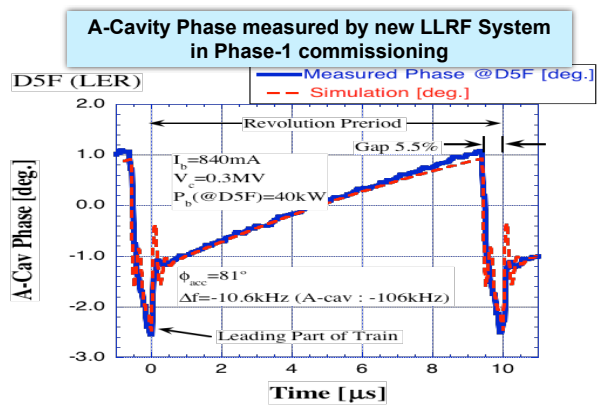

Figure 12: RF phase modulation in the accelerating cavity of ARES, which measured by new LLRF control system in Phase-1 commissioning, and the new simulation result are plotted during the revolution period.

## SUMMPARY

Phase-1 beam commissioning of SuperKEKB was successfully accomplished. Desired beam current in the two rings was achieved and sufficient vacuum scrubbing was progressed. Phase-2 is scheduled in the last quarter of JFY 2017.

Most RF system of the KEKB was reused with reinforcement for SueprKEKB. Newly developed digital LLRF control systems are applied to 9 stations at OHO section, and successfully worked in Phase-1. The new reference distribution system, in which the optical delay is stabilized by using MicroTCA system, also worked well. New LLRF control system for DR was also installed and it is ready for DR commissioning.

The $\mu$ =-1 mode damper is applied to HER, and the coupled bunch instability due to detuned cavities is suppressed successfully. Furthermore the $\mu$ =-2 and -3 mode damper sys-tem was developed for Phase-2.

The phase modulation due to the bunch gap transient was clearly observed by the new LLRF control system, and validity of the new simulation code for the evaluation of bunch gap transient effect was confirmed.